\documentclass[10pt]{article}

\usepackage{eepic}
\usepackage{epic}
\usepackage{epsfig}
\usepackage{a4wide}
\usepackage{latexsym}
\usepackage{amssymb}

\begin{document}

\def\O{{\cal O}}
\def\N{{\cal N}}
\def\>t{>_{\scriptscriptstyle{\rm T}}}
\def\enu{\epsilon_\nu}
\def\pint{\int{\d^3p\over(2\pi)^3}}
\def\gint{\int[\D g]\P[g]}
\def\nbar{\overline n}
\def\d{{\rm d}}
\def\e{{\bf e}}
\def\x{{\bf x}}
\def\y{{\bf y}}
\def\0x{\x^\smalze}
\def\sperpx{{x_\perp}}
\def\sperpk{{k_\perp}}
\def\sbperpk{{{\bf k}_\perp}}
\def\sbperpx{{{\bf x}_\perp}}
\def\perpx{{x_{\rm S}}}
\def\perpk{{k_{\rm S}}}
\def\bperpk{{{\bf k}_{\rm S}}}
\def\bperpx{{{\bf x}_{\rm S}}}
\def\p{{\bf p}}
\def\q{{\bf q}}
\def\zr{{\bf z}}
\def\R{{\bf R}}
\def\A{{\bf A}}
\def\v{{\bf v}}
\def\u{{\bf u}}
\def\w{{\bf w}}
\def\U{{\bf U}}
\def\cm{{\rm cm}}
\def\l{{\bf l}}
\def\sec{{\rm sec}}
\def\Ckol{C_{Kol}}
\def\flux{\bar\epsilon}
\def\zq{{\zeta_q}}
\def\b{b_{kpq}}
\def\bun{b^{\scriptscriptstyle (1)}_{kpq}}
\def\bdu{b^{\scriptscriptstyle (2)}_{kpq}}
\def\z0q{{\zeta^{\scriptscriptstyle{0}}_q}}
\def\smalS{{\scriptscriptstyle {\rm S}}}
\def\smalze{{\scriptscriptstyle {\rm 0}}}
\def\smalI{{\scriptscriptstyle {\rm I}}}
\def\smalII{{\scriptscriptstyle {\rm II}}}
\def\smalN{{\scriptscriptstyle {\rm N}}}
\def\smalun{{\scriptscriptstyle (1)}}
\def\smaldu{{\scriptscriptstyle (2)}}
\def\smaltr{{\scriptscriptstyle (3)}}
\def\smalL{{\scriptscriptstyle{\rm L}}}
\def\smalD{{\scriptscriptstyle{\rm D}}}
\def\smal1n{{\scriptscriptstyle (1,n)}}
\def\smaln{{\scriptscriptstyle (n)}}
\def\smalA{{\scriptscriptstyle {\rm A}}}
\def\shell{{\tt S}}
\def\ball{{\tt B}}
\def\nav{\bar N}
\def\micron{\mu{\rm m}}
\font\brm=cmr10 at 24truept
\font\bfm=cmbx10 at 15truept
\centerline{\brm A simplified model for red cell}
\centerline{\brm dynamics in small blood vessels}
\vskip 20pt
\centerline{Piero Olla}
\vskip 5pt
\centerline{ISIAtA-CNR}
\centerline{Universit\'a di Lecce}
\centerline{73100 Lecce Italy}
\vskip 20pt

\centerline{\bf Abstract}
\vskip 5pt
A simple mechanism for the confinement of red cells in the middle of narrow 
blood vessels is proposed. In the presence of a quadratic shear, red cells
deform in such a way to loose fore-aft symmetry and to achieve a fixed 
orientation with respect to the flow. This leads to a drift away from the 
vessel walls, when the vessel diameter goes below a critical value depending
on the viscoelastic properties and the dimensions of the cell.

\vskip 2cm
\vfill\eject
Blood is essentially a concentrated suspension of red cells in plasma. Other particles,
like white blood cells and platelets make up less than $1\%$ of the particulate.
Due to the high volume fraction, (around $40\%-45\%$) of the red cells, and 
to their mechanical properties, blood rheology exhibits rather complex behaviors.
Focusing on steady state situations, in particular, blood viscosity turns out to 
depend on the diameter of the vessel in which it is flowing. 

In large blood vessels, with diameters above $\sim 100\mu{\rm m}$, particles are 
randomly distributed in the plasma; the effective viscosity appears to be constant, 
and could be calculated, to a first approximation, using Einstein theory \cite{landau}. 
For diameters between $10\mu{\rm m}$ and $100\mu{\rm m}$, red cells align in the middle 
of the vessel, resulting in a decrease of the effective blood viscosity, which is 
fundamental for the well functioning of the cardiovascular system: the so called 
Fahraeus-Lindqwist effect \cite{oiknine76}. For still smaller diameters, 
corresponding to the capillary range, these cells have to squeeze their way 
through the vessel, leading to a sharp increase in the effective viscosity.
In all three cases, red cell deformability plays an important role, facilitating the 
particle flow and, as it will be shown below, contributing to the alignment process in 
the interediate range.

The alignment of red cells in the middle of small blood vessels, 
corresponds to the intuition, that particles in suspension should 
strive to minimize dissipation and to accommodate themselves in the regions 
of minimum shear. A dynamical description, however, is
not simple and requires either consideration of inertia \cite{saff65,vasseur76}, 
or of the cells non-sphericity \cite{olla97a} and deformability, and
possibly, of their mutual interactions.
\vskip 5pt
The purpose of this letter is to investigate the  hydrodynamic interactions
that stay at the basis of the Fahraeus-Lindqwist effect and to propose a simplified
model for the description of the process, based mostly on geometrical reasoning.
It appears that the drift produced by inertial corrections (see e.g. \cite{dubini95}) 
cannot account for this effect. Deformation by the quadratic part of the shear, turns 
out instead to be the essential ingredient for the migration of a red cell in a small 
blood vessel. This confirms results of numerical simulations carried out in \cite{helmy82,
leyrat94}, and, more recently, in \cite{coulliette98}. Interaction among the red cells
contributes to the effect but is not essential to the migration process. 

Consider a particle in a channel flow, which, in a reference system with origin at the 
particle center, can be written in the form: 
$$
\bar\v=\e_3(\alpha+\beta x_2+\gamma x_2^2).
\eqno(1)
$$ 
It should be mentioned that channel flow configurations are not just a device for the 
derivation of simpler models, but have physiological relevance in the description of 
blood flows in internal organs such as the lungs.
The presence of the particle produces a perturbation in the velocity field, which is
determined imposing no-slip boundary conditions at the particle surface and at the channel
walls. If the channel gap and the particle are sufficiently small, the fluid will be
described to first approximation by Stokes equation.
Thanks also to linearity of this equation, a qualitative idea of 
the velocity perturbation can be obtained, writing it as a superposition of 
images and counter-images analogous to those of electrostatics:
$ \v+\v^\smalI+\v^\smalII+... $
Here, $\v$ is the velocity perturbation in the infinite domain case, $\v^\smalI$ is the 
first image, obtained enforcing no-slip on the walls: $\v+\v^\smalI=0$, $\v^\smalII$ is
obtained imposing no-slip on the particle: $\v^\smalI+\v^\smalII=0$, and higher orders
are obtained similarly, enforcing no-slip alternatively on the walls and on the particle.

If the particle is fore-aft symmetric, expanding the velocity in vector spherical 
harmonics \cite{olla98a}, it is easy to see that
$\v^\smalN\sim a\beta R(R/l)^{2N}+b\gamma R^2(R/l)^{2N+1}$, where $R$ is the particle size 
and $l$ is the
distance to the closest wall. If the ratio $R/l$ is sufficently small, the drift of
the particle is given approximately by the value of $v^\smalI$ at the particle center.
This allows to visualize the mechanism for lift close to a wall, as shown in Fig. 1. 

\begin{figure}[hbtp]\centering
\centerline{
\psfig{figure=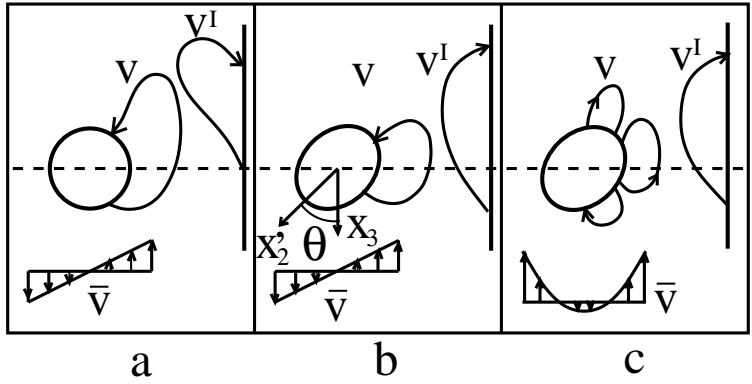,height=5.cm,angle=0.}
}
\caption{Sketch of velocity perturbation around a particle in various shear flows, and image 
velocity field produced by a wall; both velocity fields are in the particle's reference frame,
so that the wall is moving upwards. 
(a) effect of inertia: the velocity lines have an increasingly stronger tilt along the flow, 
as one gets farther away from the particle; (b) perturbation by an ellipsoidal particle in 
linear shear, in the absence of inertia; (c) the same in a quadratic shear: the velocity $\v$ 
has an octupole rather than a quadrupole symmetry, but the same drift of the previous case
occurs.}
\end{figure}

It is well known that spherical symmetry does not 
allow for transversal drift in creeping flow conditions.
Inclusion of inertia is sufficient to break the symmetry of the velocity lines (see Fig. 1a),
and this leads to a transverse migration of the order of $Re_p\beta R$ \cite{vasseur76}, 
with $Re_p={\beta R^2\over\nu}$ the particle Reynolds number and $\nu$ the kinematic 
viscosity of the fluid. Due to the smallness of $Re_p$, however, inertial drifts turn out 
to be too small to produce a Fahraeus-Linqwist effect. 

A second mechanism for symmetry breaking lies in the non-sphericity of the cells.
The drift of ellipsoidal particles has been studied in the context of 
sedimentation \cite{russel77}. More recently in \cite{olla97a,olla97b}, following 
in part \cite{keller82}, the effect of tank-treading motions
in linear shear has been taken into consideration. The velocity perturbation 
produced by an ellipsoid, is not symmetric even to $O(Re_p^0)$, and a transverse drift 
results, as illustrated in Fig. 1b-c. In the case of a linear shear, the velocity 
perturbation $\v$ decays quadratically at large distances \cite{landau}.
This leads to the expression for the drift in the presence of a wall \cite{olla97a,olla97b}:
$$
v^\smalI(l)\simeq C(\theta,R/l)\beta R\sin 2\theta,
\eqno(2)
$$
where $C(\theta,R/l)=C_0R^2/l^2+\hat cR^4/l^4+\tilde c\gamma R^4/(\beta l^3)+...$, with $C_0$ 
independent of $\theta$. The $\O((R/l)^4)$ part of the correction is basically due to the 
near field of $\v$, while the $\O(R^4/(\beta l^3))$
contribution is due to the quadratic part of the shear.  Analysis carried
out in \cite{olla98b} shows that the lift in a channel, for $R/l$ small, is obtained to a 
very good degree of confidence, by linear superposition of the drift produced individually
by each of the  walls. Thus, particles will be drifting at maximum velocity, away from
the wall, when the longest particle axis is in along the stretching direction
of $\bar v$, and towards the wall, when it is along the compressing direction. 

A tank-treading cell could maintain a fixed orientation, thus leading to a steady drift. 
Unfortunately, due to their stiffness, red cells will resist tank-treading motion in vivo 
\cite{tran84}, and will rotate as rigid objects, under the effect ot the vorticity part 
of the shear (''flipping'' motion regime).

In order for migration to take place, it is necessary that either the particle
undergoes a kind of rigid rotation which privileges orientations corresponding to
drift away rather than towards the wall, or that some new mechanism for fixed orientation
is present.  At rest, red cells can be roughly described as axisymmetric oblate
ellipsoids, with major axis $\sim 8\mu{\rm m}$ and minor axis $\sim 2\mu{\rm m}$
\cite{oiknine83}.  
Introducing a reference system $\{x'_1x'_2x'_3\}$, with $x'_1\equiv x_1$,
and $x'_3,x'_2$ respectively along the minor and the major ellipsoid axes,
the external shear can be written:
$$
\e_3x_2={1\over 2}(\e'_3x'_2-\e'_2x'_3)+
        {1\over 2}(\e'_3x'_2+\e'_2x'_3)\cos 2\theta
       +{1\over 2}(\e'_2x'_2-\e'_3x'_3)\sin 2\theta
\eqno(3)
$$
and
$$
\e_3x_2^2=\cos\theta [\e'_3(\cos^2\theta{x'_2}^2+\sin^2\theta{x'_3}^2)
-\e'_2\sin 2\theta x'_2x'_3]
$$
$$
+\sin\theta [\e'_2(\cos^2\theta{x'_3}^2+\sin^2\theta{x'_2}^2)
-\e'_3\sin 2\theta x'_2x'_3]
\eqno(4)
$$
Because of symmetry under reflection across the planes $x'_1x'_2$ and
$x'_1x'_3$, the term in $\sin 2\theta$ in Eqn. (3) does not contribute to 
torque. Turning to the quadratic shear, all terms entering Eqn. (4) 
are symmetric either under reflection: $x'_i\to -x'_i$, with $i=2,3$, or
simultaneously:  $x'_2,x'_3\to -x'_2,-x'_3$. Hence the quadratic part of 
the shear does not produce any torque on an ellipsoidal cell and does not
contribute to the angular velocity $\dot\theta$. The contribution to 
$\dot\theta$ produced by the linear part on a fore-aft symmetric cell, 
instead, has the form \cite{olla98a}: 
$$
\dot\theta=\beta (a-b\cos 2\theta),
\eqno(5)
$$
Hence, the cell will rotate with an even angular 
velocity: $\dot\theta(\theta)=\dot\theta(-\theta)$, while, from Eqn. (2),
its center will undergo a zero mean oscillatory motion in the $x_2$
direction. For an isolated cell, it is then necessary to take into 
consideration the symmetry breaking deformations produced by the external flow.

The linear shear causes a modification of the cell eccentricity,
which leads to corrections proportional to $(\Delta R/R)\sin 2\theta$, 
with $\Delta R$ the deformation, in the coefficients $b$ and $C_0$ 
$[$see Eqns. (2) and (5)$]$.
For a membrane shear elasticity $\kappa$, one has dimensionally: 
$\Delta R/R\sim\kappa^{-1}\rho\nu\beta R^3$.

The quadratic shear in Eqn. (4) is in the form $\u\cos\theta+\w\sin\theta$.
Considering a fixed cell, the two terms $\u$ and $\w$ produce respectively 
deformations which are even in $x'_2$ and $x'_3$ but break fore-aft symmetry 
in $x'_3$ and $x'_2$. These deformation will couple 
with $\w$ and $\u$ to produce a torque. Hence, for small 
deformations, this torque will be in the form: $M_1=\sin 2\theta f(\theta)$, 
with $f(\theta)$ bilinear in $\u$ and $\w$.  
From dimensional analysis, for the kind 
of flow described by Eqn. (1) and small cell deformations, one finds
$$
\dot\theta \simeq D{\rho\nu\gamma^2 R^3\over\kappa}\sin 2\theta,
\eqno(6)
$$
where $\rho$ and $\kappa$ are the plasma density and the membrane shear elasticity.
This torque has the odd dependence on $\theta$, necessary to break the symmetry
between cell orientations associated with inward and outward drifts.
The dimensionless coefficient $D$ depends on the cell rest shape and 
on the angle $\theta$, with this last dependence disappearing when the
rest shape is almost spherical \cite{olla98b}. 
Notice that the key ingredient for the production of this torque
is the violation of fore-aft symmetry. For this reason there is no 
coupling between quadratic shear and the deformation produced by the linear strain, 
and also the viceversa is true \cite{olla98b}.  

A tractable model can be obtained if one supposes that the elastic time
scale is faster than that for fluid motion, which is equivalent to assuming
small deformations. (Smallness of the deformation is necessary also to make
sure that interaction with the walls are the only source of drifts in the problem
\cite{olla98a}).

Consider then a channel flow of width $2d$ and fluid velocity 
at the center $V$ (see Fig. 2). After expressing all lengths and times in units 
of $d$ and $d/V$, the model equations will read:
$$
\left\{\begin{array}{l}
\dot x=2r^3x(C(\theta,1-x)-C(\theta,1+x))\sin 2\theta
\\
\dot\theta=2 x(a-b\cos 2\theta)+Dr^3V\rho\nu\kappa^{-1}\sin 2\theta
\end{array}\right.
\eqno(7)
$$
where $r=R/d$ and $x=x_2/d$ are the dimensionless particle radius and distance from 
the vessel axis. Following Eqn. (2), the two terms $C$ in the first line of Eqn. (7) 
account for the transverse drift, as sum of the individual contributions from the two 
walls. The factor $x$ in front gives the strength of the linear part of the shear at 
the particle center. The second line contains the balance between the torque from the 
vorticity part of $\bar\v$ (even in $\theta$) and the counter-torque (odd in $\theta$) 
provided by the interaction between quadratic shear and the cell deformation.

Taking $D$ independent of $\theta$, as in the spheroidal limit, allows to 
analyze the dynamics of the system (actually, it is enough that $D$ does not change sign). 
Its phase portrait is shown in Fig. 3 and clearly indicates the presence of a stable fixed 
point at $x=0,\theta=({1\over 2}+n)\pi$
corresponding to red cell alignment in the channel axis, perpendicular to the flow.
\begin{figure}[hbtp]\centering
\centerline{
\psfig{figure=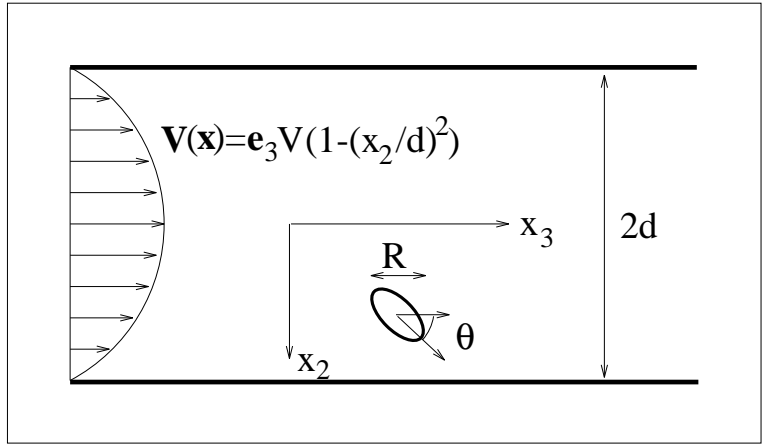,height=10.cm,angle=-90.}
}
\caption{Ellipsoidal cell in a channel flow. At the particular position and orientation
represented, the cell is moving downwards while rotating clockwise (i.e. $\dot\theta<0$).}
\end{figure}
However, if $r$ is small, the approach to the channel axis will be very slow,
as it is easy to see from the following considerations. The first of Eqn. (8), for $r$ small 
and away from the wall, can be approximated as $\dot x\sim r^3x^2\sin 2\theta$;
substituting into the second one and expanding in powers of $r$ one finds:
$$
{\d\ln x\over \d\theta}\sim{r^3\sin 2\theta\over a-b\cos 2\theta}
\Big(1-{D r^3\sin 2\theta\over 2x(a-b\cos 2\theta)}+...\Big)
\eqno(8)
$$
and the drift will be $\O(r^6)$. Terms of the same order are produced including the 
corrections to $b$ and $C_0$ produced by the deformations from the linear shear.
Thus, the fixed point basin of attraction has an 
effective boundary at $\hat x\sim r^3V\rho\nu\kappa^{-1}$, the width of the region of $x$
occupied by the curve of turning points for $\theta$ (the dotted line in Fig. 3). An
''isolated particle'' Fahraeus-Lindqwist effect will be present only when the width 
of this region and the one of the channel are comparable, and this leads to the condition on
$d$:
$$
d<E\Big({V\rho\nu\over\kappa}\Big)^{1\over 3}R,
\eqno(9)
$$
with $E$ depending on the rest shape of the cell through the coefficients $a$, $b$,
$D$ and $C_0$.
\begin{figure}[hbtp]\centering
\centerline{
\epsfig{file=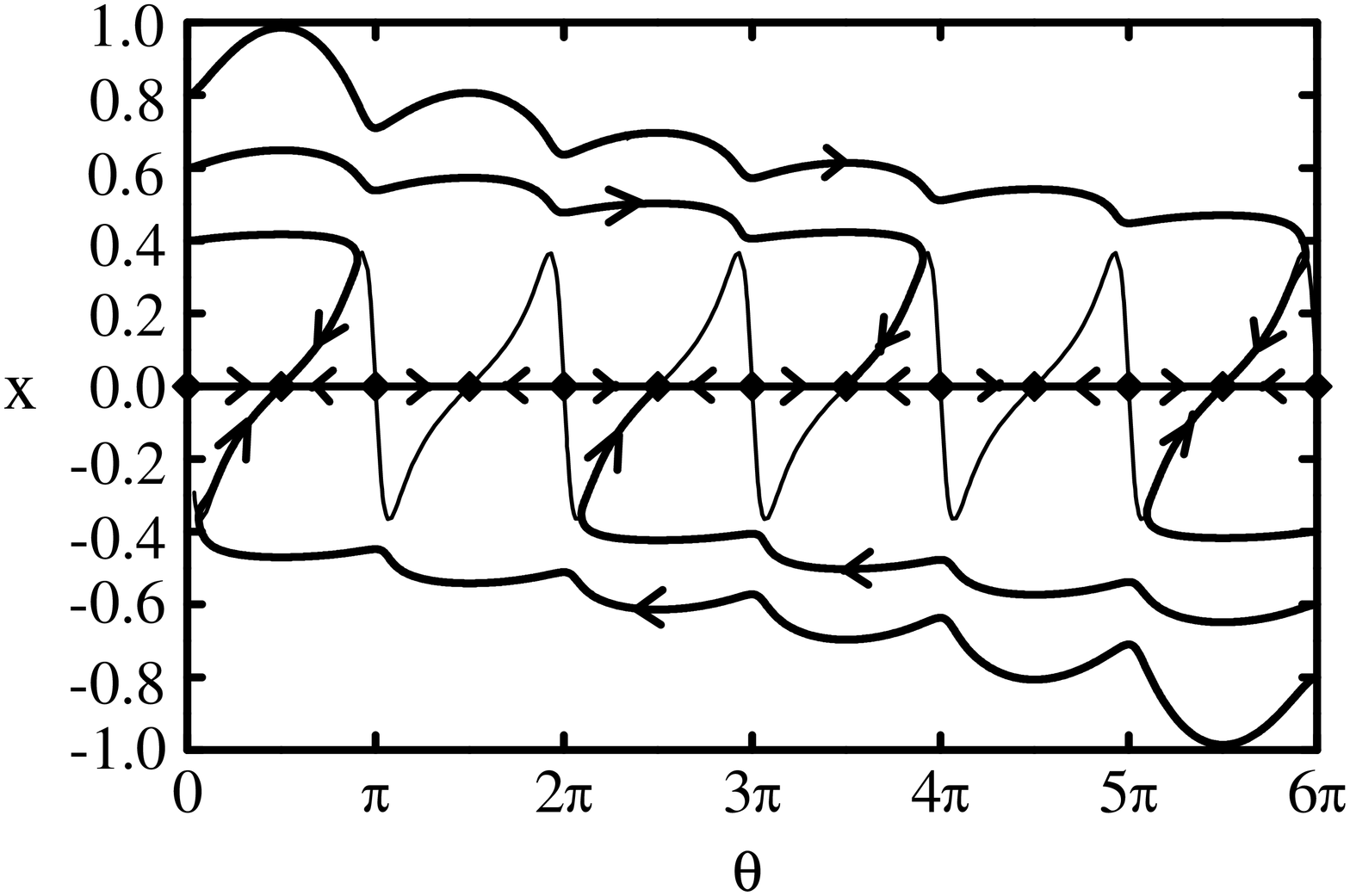,height=9.cm,angle=-90.}
}
\psfull
\caption{Phase diagram of the trajectories in the $\theta-x$ plane for $r=0.2$. The dashed
line indicates the curve $\dot\theta=0$ of turning points for $\theta(t)$. The trajectory 
pattern illustrated depends solely on where the line of turning points crosses the
$x$-axis and on where the trajectories bend towards or away from the wall, which
could be modified only if $C$ presented sign changes.
}
\end{figure}
The phase diagram in Fig. 3 is obtained for values of the parameters: $C_0=0.5$,
$a=0.5$, $b=0.4$, and $DV\rho\nu\kappa^{-1}=20$. The membrane shear elasticity is 
$\kappa\simeq 3\times 10^{-3}{\rm dyne}\,{\rm cm}^{-1}$ \cite{drochon90},
while the plasma density is the same of water: $\rho\simeq 1{\rm g}\,{\rm cm}^{-3}$;
for a flow velocity of the order of a centimeter per second, this corresponds to a value $D\sim 6$
in Eqn. (7). The drift $C(r,z)$ has been modelled putting: $C(r,z)=C_0(r^2+z^2)^{-1}$. 
It is important to stress, however, that the precise form of $C$ is unimportant, provided
it does not change sign, which, if occurring in correspondence of the line
of turning points for $\dot\theta$, could lead to spurious fixed points at
$(\theta,x)\ne ({\pi\over 2}+n\pi,0)$. In the present case (see Fig. 3), the distance
of the line of turning points from the walls, however, is sufficient to guarantee that the 
far field expression
for $C(\theta,x)$ has at least qualitative value, and that changes of sign in $C$ should not
be a problem.

In the situation considered, one finds the transition to a Fahraeus-Linqdwist effect dominated 
situation for $d\lesssim 5R$. The particular value of $d$, however is not interesting in
itself, depending on the choice of the model parameters. What is interesting, is the 
sharpness of the transition to the Fahraeus-Lindqwist dominated regime, which is a consequence
of the strong dependence of the drift on the parameter $r$ $[$see Eqn. (8)$]$.
A red cell starting at $(\theta,x)=(0,0.7)$, while drifting to $x=r$, will
travel a distance $L$ along the flow, that jumps from $L\simeq 30 d$, for $r=.25$, to 
$L\simeq 1500d$ for $r=.15$.  
\vskip 5pt
To summarize, a qualitative picture arises, of a flow in a small blood vessel,
separated into an inner and outer region, dominated respectively by the 
quadratic and the linear part of the shear. An isolated red cell in the external 
region, will remain there, in a state of flipping motion for a very long time. Once in 
the internal region, instead, it will dispose itself perpendicular to the flow, assuming,
under the effect of the quadratic shear, a parachute kind of shape; at the same 
time, it will be pushed towards the vessel axis by interaction with the walls.
A Fahraeus-Lindqwist effect will therefore occur, if enough of the vessel interior lies 
in this quadratic shear dominated region. 

In the present study, only the behavior of an isolated red cell has been taken into exam. 
Consideration of the mutual interaction of red cells in real blood, however, 
is going to strengthen the results obtained. Inter-particle interactions will act like a noise 
in Eqn. (8) and are going to 
produce two effects on the dynamics of a single red cell: the destabilization
of the fixed points at $(\theta,x)=({\pi\over 2}+n\pi,0)$ and the transverse diffusion 
of trajectories in the flipping motion region.
If the quadratic region is wide enough, say, $\hat x\sim r^3V\rho\nu\kappa^{-1}>r$,
the second effect will dominate, speeding up collapse of the trajectories onto the channel axis.
Thus, a larger values for the value of $d$ giving the transition to the 
Fahraeus-Lindqwist effect dominated regime, is going to be expected, with a dependence on
the flow parameters given by: $d\lesssim\Big({V\rho\nu/\kappa}\Big)^{1\over 2}R$, in
place of the analogous expression provided by Eqn. (9).

\vskip 10pt
\noindent{\bf Aknowledgements}: I would like to thank Massimiliano Tuveri, 
Howard Stone, Dominique Barth\'es-Biesel and Claudio Tebaldi for interesting and helpful 
conversation. Part of this research was carried on at CRS4 and at the 
Laboratoire de Mod\'elisation en M\'ecanique in Jussieu.
I would like to thank Gianluigi Zanetti and Stephane Zaleski for hospitality.


\vskip 20pt
\begin{thebibliography}{99}
\bibitem{landau} L.D. Landau and E.M. Lifshitz, {\it Fluid Mechanics}, 
(Pergamon Press, Oxford, 1982)
\bibitem{oiknine76} F. Azelvadre and C. Oiknine, Biorheology {\bf 13}, 315 (1976) 
\bibitem{saff65} P.G. Saffmann, J. Fluid Mech. {\bf 22}, 385 (1965)
\bibitem{vasseur76} P. Vasseur and R.G. Cox, J. Fluid Mech. {\bf 78}, 385 (1976)
\bibitem{olla97a} P. Olla, J. Phys. A (Math. Gen.) {\bf 30}, 317 (1997)
\bibitem{dubini95} G. Dubini, R. Pietrabissa and F.M. Montevecchi, Med. Eng. Phys.
{\bf 17} 609 (1995)
\bibitem{helmy82} A. Helmy and D. Barthes-Biesel, J. M\'ec. Th\'eor. Appl. {\bf 1}, 859 (1982)
\bibitem{leyrat94} A. Leyrat and D. Barthes-Biesel, J. Fluid Mech. {\bf 279}, 135 (1994)
\bibitem{coulliette98} C. Coulliette and C. Pozrikidis, J. Fluid Mech. {\bf 258}, 1 (1998)
\bibitem{olla98a} P. Olla, (1998) Submitted to J. Fluid Mech. Chao-Dyn/9808022
\bibitem{russel77} W.B. Russel et Al., J. Fluid Mech. {\bf 83}, 273 (1977)
\bibitem{olla97b} P. Olla, J. Phys. II (France) {\bf 7}, 153 (1997)
\bibitem{keller82} S.R. Keller and R. Skalak, J. Fluid Mech. {\bf 120}, 27 (1982)
\bibitem{tran84} R. Tran-Son-Tay, S.P Sutera and P.R. Rao, J. Biophys. Soc. {\bf 46}, 65 (1984)
\bibitem{oiknine83} C. Oiknine, in {\it Cardiovascular Engineering}, Ed. D.N Ghista, 
E. Van Vollenhoven, W.-J. Yang, H. Reul and W. Bleifeld (Karger, Basel, 1983), pp. 2-25
\bibitem{olla98b} P. Olla (1998) To be submitted \
\bibitem{drochon90} A. Drochon et Al. J. Biomech. Engn.  {\bf 112}, 241 (1990)
\end{thebibliography}
\end{document}